\documentclass[letterpaper]{article}
\usepackage{iccc}

\usepackage{graphicx}
\usepackage{times}
\usepackage{helvet}
\usepackage{courier}
\usepackage{fancyhdr}
\usepackage{amsfonts}
\usepackage{url}
\fancypagestyle{firstpage}{
  \fancyhf{}
  \fancyhead[C]{\small Accepted at the 17th International Conference on Computational Creativity (ICCC 2026)}
  
}
\title{Seeing Differently: Modeling Interpretive Perspectives in Computational Creativity using a Four-World Framework\\
Paper type: Study Paper}

\author{Prerna Luthra\\
Independent Researcher\\
prerna@samvedna.ai\\
}
\begin{document} 
\maketitle
\thispagestyle{firstpage}
\begin{abstract}
\begin{quote}
Creativity in computational systems is often evaluated as an objective property of artifacts, with existing Computational Creativity (CC) frameworks assessing creative merit at the level of outputs or systems rather than interpretive context. However, artistic meaning is inherently perspective-dependent and can vary across viewers and critical traditions. This paper proposes a computational approach to modeling interpretive perspectives rather than treating creativity as a single measurable construct. The study adopts a twelve-trait creativity framework, organized across four conceptual domains, and operationalizes it through three evaluative personas: formalist, social-historical, and iconographic. Using 1,069 artworks from the SemArt dataset, the analysis generates 38,484 persona-based evaluations to examine how perspectives shape creativity assessment. Results show systematic divergence across perspectives, with traits such as Social Reflexivity exhibiting strong viewpoint sensitivity. Linear probing of CLIP image embeddings reveals that perspectives correspond to distinct orientation vectors in representation space, suggesting that creativity evaluation depends on which visual features become salient under each persona condition. These findings support a relational view of creativity and indicate that incorporating multiple evaluative perspectives could enable co-creative systems to support interpretively diverse human collaborators.
\end{quote}
\end{abstract}

\section{Introduction}
Computational approaches to creativity evaluation have made substantial progress in operationalizing properties such as novelty, value, and surprise as measurable attributes of artifacts or systems ~\cite{boden2004,ritchie2007}. These approaches typically assume that creative merit can be assessed objectively, that a sufficiently well-designed rubric, applied consistently, will yield stable evaluations across observers. However, this assumption sits in tension with a longstanding insight from aesthetics and art history: the meaning of a creative work is not fixed, but emerges through the relationship between the artifact and the interpretive perspective brought to bear on it ~\cite{panofsky1955,eco1976}.

This tension raises an important question for computational creativity: if human evaluators applying different analytical frameworks systematically disagree about the creative properties of the same artifact, should computational evaluation systems model this diversity rather than collapse it into a single metric?

This paper investigates that question empirically. We adopt a twelve-trait creativity framework ~\cite{anonymous2025fourworld} organized across four conceptual domains---Inner, Outer, Imaginative, and Moral, and operationalize three interpretive perspectives drawn from established art-critical traditions: Formalist, Social–Historical, and Iconographic. Using GPT-4.1 as a vision–language evaluator conditioned on persona prompts, we evaluate 1,069 artworks from the SemArt dataset under each perspective and analyze how the resulting trait scores diverge. We further train ridge regression probes on CLIP image embeddings to examine whether different perspectives correspond to distinct orientations within a shared visual representation space.

Our results show that creativity assessments vary systematically across perspectives, that certain traits are substantially more perspective-sensitive than others, and that different evaluative lenses are associated with different orientations in representation space. Taken together, these findings support a relational view of creativity and suggest that modeling interpretive plurality is both computationally tractable and theoretically motivated.

The relational framing of creativity adopted here has substantial precedent in creativity research. Rhodes's classical four Ps (person, process, product, press) already names press---the receiving environment---as a component of creative assessment ~\cite{rhodes1961}, and Jordanous’s PPPPerspectives model extends this by treating press/environment as central to evaluative judgment~\cite{jordanous2016four}. Glăveanu's Five A's framework recasts creativity as a distributed phenomenon involving actor, artifact, audience, action, and affordances~\cite{glaveanu2013fiveas}; Kantosalo and Takala's Five C's framework foregrounds the receiving community within co-creative systems~\cite{kantosalo2020five}; and Sternberg and Karami's 8P framework integrates these threads across product, process, and perception-oriented accounts~\cite{sternberg2021}. What these frameworks share---and what the present work builds on---is a recognition that creativity evaluation depends on more than the artifact in isolation. Our contribution is to operationalize one specific dimension of this relational picture: the evaluative perspective brought to bear on the artifact, studied empirically across a large set of artworks.

\section{Four-World Creativity Framework}
This study adopts a Four-World framework for conceptualizing creativity across four complementary domains: the Inner World, Outer World, Imaginative World, and Moral World \cite{anonymous2025fourworld}. The framework was developed by synthesizing insights from psychological theories of creativity and semi-structured interviews with practicing artists, with the goal of capturing dimensions of creative meaning that extend beyond novelty or technical execution. Table 1 illustrates all traits associated with the four worlds.
\begin{table}[t]
\centering
\small
\setlength{\tabcolsep}{4pt}
\caption{Four-World creativity framework and associated traits.}
\begin{tabular}{lp{5cm}}
\hline
\textbf{World} & \textbf{Traits} \\
\hline
Inner World & Emotional Intensity; Memory Imprint; Personal Symbolism \\
Outer World & Cultural Situatedness; Environmental Dialogicity; Social Reflexivity \\
Imaginative World & Surreal Divergence; Symbolic Density; Playful Subversion \\
Moral World & Ethical Provocation; Collective Resonance; Redemptive Arc \\
\hline
\end{tabular}
\label{tab:four_worlds}
\end{table}

\textbf{Inner World} represents introspective dimensions of creativity grounded in emotional experience, autobiographical memory, and personal symbolic systems. Works within this domain often convey subjective feeling, internal narrative, or idiosyncratic symbolic language reflecting the creator’s inner life. 

\textbf{Outer World} situates creative works within cultural, environmental, and social contexts. Here creativity emerges through engagement with historical traditions, relationships between humans and their environments, and dialogue with social audiences and institutions.

\textbf{Imaginative World} captures creativity as the transformation of reality through symbolic invention, surreal imagery, and playful experimentation. These traits reflect the ability of creative works to construct alternate logics, layered symbolism, or unexpected conceptual combinations.

\textbf{Moral World} addresses the ethical and communal dimensions of creativity. Works in this domain may provoke reflection on moral dilemmas, resonate with shared collective experiences, or depict narratives of transformation, justice, and renewal.

Across these four domains the framework defines twelve traits that operationalize these dimensions of creative expression. In the present study, we adopt this framework not as a predictive model of creativity but as a structured vocabulary for examining how different interpretive perspectives evaluate the same artwork. Rather than assuming that creativity traits are intrinsic properties of artifacts, we investigate how these traits may be interpreted differently depending on the evaluative viewpoint applied. Note that the twelve-trait rubric and scoring prompts follow the formulation specified in \cite{anonymous2025fourworld}. For completeness, the exact prompts used in this study are provided in Appendix A.

A useful distinction within the creativity literature is between \emph{meta-frameworks}, which name the constituent components of a complete account of creativity (e.g., actor, artifact, audience, process, environment), and \emph{rubrics}, which score individual artifacts on measurable dimensions.
The relational frameworks discussed above (Glăveanu's Five A's, Jordanous's PPPPerspectives, Kantosalo and Takala's Five C's, Sternberg and Karami's 8P) are meta-frameworks: they establish that a complete creativity account must address the receiver and the context, but they do not themselves provide a trait-level rubric for scoring works. The artifact-evaluation rubrics that do exist: Jordanous's fourteen creativity components~\cite{jordanous2012evaluating}, Ritchie's empirical criteria~\cite{ritchie2007}, and Boden's H/P-creativity distinctions~\cite{boden2004} were developed to evaluate creative \emph{systems} or to make high-level judgments of novelty, value, and surprise. Several of their dimensions (e.g., Active Involvement and Persistence, Domain Competence, Intention) are properties of the creator rather than features scoreable from a finished work; others operate at too coarse a granularity to localize where different interpretive perspectives disagree.

The present study sits at the intersection of these two strands. It takes from the relational frameworks the commitment that creativity evaluation depends on the perspective applied, and it requires a rubric that (i) scores at the artifact level, (ii) covers \emph{semantic} dimensions where interpretive traditions are likely to diverge, and (iii) permits perspective effects to be localized to specific dimensions of meaning rather than collapsed into a single creativity judgment. Existing rubrics were designed for different purposes: characterizing creative systems or making global evaluative judgments of novelty, value, and surprise, rather than localizing semantic differences in how artworks are interpreted. 

The Four-World framework was originally developed for analyzing visual creativity across psychological, social, imaginative, and ethical dimensions, synthesizing psychological theories of creativity with semi-structured interviews with practicing artists~\cite{anonymous2025fourworld}. We adopt it for the present study because its twelve traits span precisely the kinds of meaning-making dimensions---introspective, contextual, imaginative, and ethical---where formalist, social-historical, and iconographic readings diverge systematically, as shown in the results that follow. The framework functions here as an instrument for examining interpretive divergence, not as a claim that creativity reduces to trait scores.

\section{Modeling Interpretive Perspectives}
While the Four-World framework defines dimensions of creative expression, the interpretation of these dimensions may vary depending on the evaluative perspective applied. Art historical and aesthetic traditions have long emphasized that artworks can be interpreted through different analytical lenses, each prioritizing different forms of evidence and meaning ~\cite{panofsky1955,eco1976}. Rather than assuming that creativity traits are intrinsic properties of artifacts, this study models interpretive perspectives as structured evaluative viewpoints.
We operationalize interpretive perspectives using three personas derived from established art-critical traditions. These perspectives were chosen because they represent widely used analytical approaches in art history and are well suited to analyzing artworks in the SemArt dataset used in this study, which primarily contains European fine-art paintings:

\textbf{Formalist} perspective, which emphasizes visual composition, form, color, and stylistic execution while largely ignoring historical or social context, consistent with formalist approaches to art criticism ~\cite{greenberg1961}.

\textbf{Social–historical} perspective, which interprets artworks through their cultural, political, and historical conditions, focusing on how creative works reflect or critique social structures ~\cite{hauser1951,clark1984}.

\textbf{Iconographic} perspective, which focuses on symbolic content, narrative meaning, and the interpretation of visual motifs and allegorical elements following iconographic traditions in art history ~\cite{panofsky1955}.

These perspectives are implemented through persona-conditioned prompts that instruct a vision–language model to evaluate artworks according to a particular interpretive lens. Each persona prompt directs the model to apply the same twelve-trait rubric derived from the Four-World framework but may weight different forms of visual or contextual evidence when assigning scores. The prompts explicitly frame the evaluation as an interpretive perspective rather than an objective judgment, encouraging the model to apply the lens consistently when scoring traits. Full persona prompts are provided in Appendix A.

\section{Experiment}

\paragraph{Dataset}
We use the SemArt dataset ~\cite{garcia2018semart}, a corpus of approximately 21,000 European fine-art paintings designed for semantic art understanding, spanning multiple historical periods, styles, and artistic schools. We analyze 1,069 artworks from the standardized test partition. Each artwork includes metadata fields — title, artist, date, technique, artistic school, and a textual description — which are provided to the model alongside the image during evaluation. SemArt's focus on European fine-art traditions makes it well suited for examining interpretive perspectives derived from established art-historical analysis traditions.

\paragraph{Persona-Conditioned Trait Scoring}
To operationalize interpretive perspectives, we evaluate each artwork using a vision-language model GPT-4.1 ~\cite{openai2025gpt41} conditioned on persona prompts corresponding to three art-critical traditions: Formalist, Social–Historical, and Iconographic.
Each persona receives the same input: the artwork image, artwork metadata (title, author, date, technique, school), the textual description from the dataset, and the twelve-trait rubric derived from the Four-World framework.
The model is instructed to assign a 0–4 score for each trait (0 = absent, 4 = very strong) and return a structured JSON response containing: trait name, associated world, score, short reasoning.
Evaluation is performed with $temperature = 0$ and $top\_p = 1$ to minimize output variability. Each artwork is scored independently by all three personas.
Across 1,069 artworks and three personas, this process produces 38,484 trait scores (12 traits × 3 personas × 1,069 images). These scores form the basis for subsequent analyses of cross-perspective divergence.

\paragraph{Visual Embeddings}
To analyze how interpretive perspectives relate to visual representation, each artwork is encoded using CLIP ViT-B/32, producing a 512-dimensional image embedding ~\cite{radford2021clip}.
CLIP image embeddings are computed once per image and reused across analyses. Images are preprocessed using the model’s standard preprocessing pipeline and encoded using the image encoder.

\paragraph{Linear Probing of Interpretive Orientation}
To examine how different perspectives interpret the same trait within a shared visual representation space, we train ridge regression probes that map CLIP embeddings to persona-specific trait scores ~\cite{alain2017probes}. For each trait $T$ and persona $P$, we solve

\[
\min_{w_{T,P}} \|X w_{T,P} - y_{T,P}\|^2 + \lambda \|w_{T,P}\|^2
\]

where $X \in \mathbb{R}^{N \times 512}$ are standardized CLIP embeddings for $N$ images, $y_{T,P}$ are the trait scores assigned by persona $P$, and $\lambda$ is the regularization strength selected via cross-validation. The resulting weight vector $w_{T,P}$ defines a direction in embedding space capturing how that perspective associates visual features with the trait.

To compare perspectives, we compute cosine similarity between persona vectors for the same trait. Lower similarity indicates that the persona-conditioned probes weight different visual features when predicting the trait. Across twelve traits and three personas, this procedure yields 36 orientation vectors that enable quantitative analysis of interpretive divergence in representation space.

\section{Results}

\begin{figure}[t]
\centering
\includegraphics[width=\columnwidth]{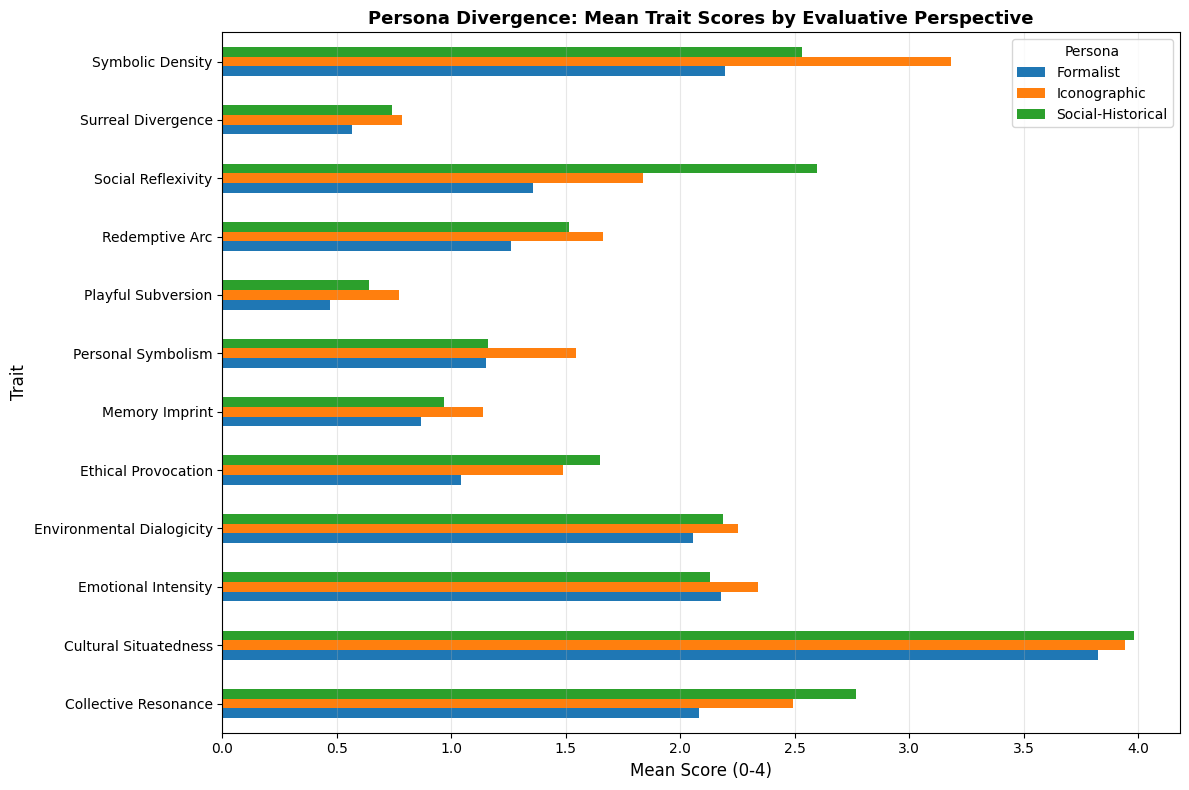}
\caption{Mean trait scores by evaluative persona across the twelve creativity traits. The Iconographic persona elevates symbolic traits; the Social--Historical persona elevates socially oriented traits; the Formalist persona consistently scores contextual traits lower.}
\label{fig:persona_analysis}
\end{figure}

\paragraph{Persona Divergence}
To examine whether interpretive perspectives systematically influence creativity evaluations, we compare mean trait scores assigned by each persona across the 1{,}069 artworks in the SemArt test set. Figure~\ref{fig:persona_analysis} shows the average score (0--4 scale) for each of the twelve traits under the Formalist, Social--Historical, and Iconographic perspectives; 95\% bootstrap confidence intervals (2{,}000 resamples) for all reported means are given in Table~\ref{tab:stats}.

Because trait scores are ordinal (0--4) and evaluated repeatedly across the same artworks, we use nonparametric repeated-measures tests. For each trait, we apply a Friedman rank-sum test treating the three personas as repeated measures over the same 1{,}069 artworks. The omnibus Friedman test is significant for every trait (all $p < 10^{-49}$), indicating systematic differences in persona-conditioned evaluations. We then apply pairwise Wilcoxon signed-rank tests across personas with Holm--Bonferroni correction. Given the large sample size, we focus primarily on Kendall's coefficient of concordance $W$ as an effect-size measure. Larger $W$ values indicate stronger and more consistent persona-conditioned differences in scoring across artworks.

\begin{table*}[t]
\centering
\small
\setlength{\tabcolsep}{5pt}
\caption{Persona-conditioned trait scores with 95\% bootstrap confidence intervals, Friedman omnibus test, Kendall's $W$ effect size, and mean cross-persona variance. All Friedman tests are significant at $p < 10^{-49}$ ($n = 1{,}069$). Traits are ordered by Kendall's $W$. All pairwise Wilcoxon signed-rank tests are significant after Holm--Bonferroni correction except Personal Symbolism (Formalist vs.\ Social--Historical, $p = 0.35$), marked with $\dagger$.}
\label{tab:stats}
\begin{tabular}{lcccccc}
\hline
\textbf{Trait} & \textbf{Formalist} & \textbf{Iconographic} & \textbf{Social--Hist.} & \textbf{Friedman $\chi^2$} & \textbf{Kendall's $W$} & \textbf{Mean Var.} \\
\hline
Social Reflexivity            & 1.35 [1.31, 1.40] & 1.84 [1.80, 1.88] & 2.60 [2.56, 2.63] & 1680.83 & 0.786 & 0.354 \\
Symbolic Density              & 2.20 [2.15, 2.25] & 3.18 [3.13, 3.23] & 2.53 [2.48, 2.58] & 1521.29 & 0.712 & 0.244 \\
Collective Resonance          & 2.08 [2.03, 2.14] & 2.49 [2.44, 2.55] & 2.77 [2.71, 2.82] & 1021.08 & 0.478 & 0.169 \\
Ethical Provocation           & 1.04 [0.98, 1.11] & 1.49 [1.42, 1.56] & 1.65 [1.58, 1.71] &  879.96 & 0.412 & 0.164 \\
Personal Symbolism$^\dagger$  & 1.15 [1.12, 1.19] & 1.54 [1.51, 1.59] & 1.16 [1.12, 1.20] &  674.99 & 0.316 & 0.107 \\
Redemptive Arc                & 1.26 [1.19, 1.33] & 1.66 [1.59, 1.74] & 1.52 [1.44, 1.59] &  522.49 & 0.244 & 0.114 \\
Playful Subversion            & 0.47 [0.43, 0.51] & 0.77 [0.72, 0.82] & 0.64 [0.60, 0.69] &  396.61 & 0.186 & 0.083 \\
Memory Imprint                & 0.87 [0.83, 0.91] & 1.14 [1.10, 1.18] & 0.97 [0.93, 1.01] &  361.74 & 0.169 & 0.075 \\
Emotional Intensity           & 2.18 [2.13, 2.23] & 2.34 [2.29, 2.39] & 2.13 [2.08, 2.19] &  331.64 & 0.155 & 0.051 \\
Surreal Divergence            & 0.57 [0.52, 0.62] & 0.78 [0.73, 0.84] & 0.74 [0.69, 0.80] &  300.69 & 0.141 & 0.064 \\
Cultural Situatedness         & 3.83 [3.80, 3.85] & 3.94 [3.92, 3.96] & 3.98 [3.97, 3.99] &  257.38 & 0.120 & 0.038 \\
Environmental Dialogicity     & 2.06 [2.00, 2.11] & 2.25 [2.19, 2.31] & 2.19 [2.13, 2.24] &  224.49 & 0.105 & 0.062 \\
\hline
\end{tabular}
\end{table*}

Several patterns emerge. The Social--Historical perspective assigns substantially higher scores to socially oriented traits. Social Reflexivity receives a mean of 2.60 [95\% CI: 2.56, 2.63] under the Social--Historical persona, compared to 1.35 [1.31, 1.40] under the Formalist persona---the largest persona effect in the dataset (Kendall's $W = 0.79$; all three pairwise comparisons $p < 10^{-100}$ after correction). Collective Resonance shows a similar pattern, with the Social--Historical persona scoring 2.77 [2.71, 2.82] versus the Formalist 2.08 [2.03, 2.14] ($W = 0.48$). These differences are consistent with social art history's emphasis on artworks as expressions of cultural structures, political contexts, and collective meaning.

The Iconographic perspective assigns notably higher scores to symbolic traits. Symbolic Density receives a mean of 3.18 [3.13, 3.23] under the Iconographic persona, compared with 2.20 [2.15, 2.25] under the Formalist persona and 2.53 [2.48, 2.58] under the Social--Historical persona (Kendall's $W = 0.71$; all pairwise comparisons $p < 10^{-77}$ after correction). The pattern aligns with iconographic traditions that interpret artworks primarily through symbolic motifs and narrative meaning.

In contrast, the Formalist perspective assigns lower scores to traits grounded in contextual interpretation, including Ethical Provocation ($W = 0.41$), Collective Resonance ($W = 0.48$), and Social Reflexivity. Because the Formalist prompt directs the model toward visual form and compositional structure rather than contextual interpretation, traits related to social or symbolic meaning receive lower scores under this condition.

The pairwise tests also reveal informative nuance. For Personal Symbolism, the Formalist and Social--Historical personas produce essentially identical means (1.15 [1.12, 1.19] and 1.16 [1.12, 1.20]; $p = 0.35$ after correction)---the only non-significant pairwise comparison in the dataset---while the Iconographic persona scores substantially higher (1.54 [1.51, 1.59]). The trait is therefore not uniformly perspective-sensitive: only the Iconographic lens systematically diverges from the other two, consistent with its explicit attention to symbolic content. Notably, the orientation analysis below shows that the Iconographic probe weights substantially different visual features for Personal Symbolism than the other two probes (cosine similarities of 0.51 with Formalist and 0.53 with Social--Historical, versus 0.70 between Formalist and Social--Historical themselves). The Iconographic lens thus diverges from the other two at both the score level and the feature-weighting level, while the Formalist and Social--Historical probes converge on both---a coherent pattern in which one perspective stands genuinely apart.

Some traits show small persona effects despite statistically significant Friedman tests. Cultural Situatedness ranges narrowly from 3.83 to 3.98 across personas ($W = 0.12$), Environmental Dialogicity from 2.06 to 2.25 ($W = 0.11$), and Emotional Intensity from 2.13 to 2.34 ($W = 0.16$). These traits appear less sensitive to interpretive framing, suggesting that certain dimensions of creative expression are more robust across evaluative viewpoints.

Memory Imprint (0.87--1.14) and Playful Subversion (0.47--0.77) are scored low by all three personas within the SemArt corpus. We are cautious about interpreting this pattern: low scores may reflect a genuine rarity of autobiographical or experimentally playful dimensions in the artworks in this dataset, but they may also reflect what is recoverable from a static image together with short textual metadata---autobiographical memory is often inaccessible without external biographical context, and stylistic playfulness may not register through the cues available in this study. We therefore do not extrapolate these patterns to European fine-art traditions more broadly.

Overall, these results indicate that creativity assessments vary systematically with the interpretive perspective applied, with the magnitude of variation differing substantially across traits.

\paragraph{Trait Sensitivity Across Perspectives}
To quantify which creativity dimensions are most dependent on perspective, we examine two complementary measures. The first is mean cross-persona variance: for each artwork, the variance of the three persona scores is computed for a given trait and then averaged across all 1{,}069 images. The second is Kendall's coefficient of concordance $W$ from the Friedman test described above, which serves as a non-parametric effect-size measure of how strongly persona conditioning rank-orders scoring on a trait. Both measures are reported in Table~\ref{tab:stats}.

The most perspective-sensitive trait is \textit{Social Reflexivity} (mean variance $= 0.354$; Kendall's $W = 0.79$), indicating substantial disagreement across perspectives about whether an artwork engages with social structures or collective issues. Other traits with relatively high persona dependence include \textit{Symbolic Density} ($W = 0.71$), \textit{Collective Resonance} ($W = 0.48$), and \textit{Ethical Provocation} ($W = 0.41$). These traits involve interpretive judgments that extend beyond purely visual properties, which may explain their greater sensitivity to perspective.

In contrast, several traits exhibit low persona dependence. \textit{Environmental Dialogicity} ($W = 0.11$), \textit{Cultural Situatedness} ($W = 0.12$), and \textit{Surreal Divergence} ($W = 0.14$) show comparatively stable scores across persona conditions despite statistically significant Friedman tests; the statistical significance at $n = 1{,}069$ reflects the high power of the design rather than a substantively large persona effect. The small effect sizes indicate that these traits rely more directly on observable visual cues or shared interpretive conventions.

The two measures yield closely aligned rankings: the top four traits are identical under both orderings, and the same five low-effect traits (Cultural Situatedness, Environmental Dialogicity, Emotional Intensity, Surreal Divergence, and Memory Imprint) cluster at the bottom of both, though with minor differences in exact order. Together they suggest that creativity traits occupy different positions along a spectrum between interpretive dependence and perceptual stability, a pattern we examine further by analyzing whether perspective-sensitive traits also correspond to divergent orientations in visual representation space.

\paragraph{Boundary Objects}

Star and Griesemer's concept of boundary objects describes artifacts that hold distinct meanings across interpretive communities while retaining a shared identity across them~\cite{star1989boundary}. We adopt this notion in a limited, analogical sense: rather than studying communities of human interpreters, we use persona-conditioned evaluations as structured proxies for distinct interpretive traditions in art criticism. Artworks producing the highest cross-persona variance can then be examined as candidate boundary cases---works that admit substantially different readings depending on the analytical lens applied. For each artwork, we compute the mean variance of trait scores across personas; artworks with the highest disagreement scores serve as candidate boundary cases.

The most divergent example is \textit{Water Glass and Jug} (c.~1760) by Jean-Baptiste-Siméon Chardin, a still-life painting depicting a glass of water, an earthenware jug, and several garlic bulbs arranged on a simple surface as shown in Figure 2. A detailed breakdown of its scores illustrates how different perspectives attribute distinct meanings to the same artifact.

\begin{figure}[t]
\centering
\includegraphics[width=0.8\columnwidth]{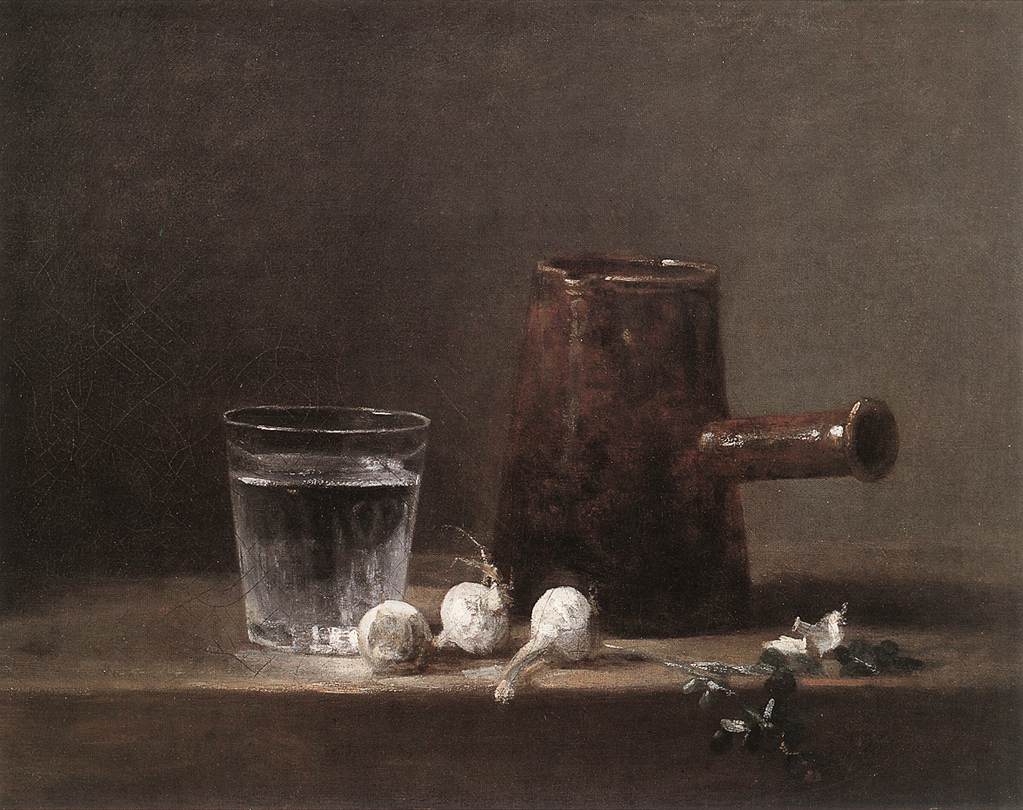}

\vspace{4pt}

\small
\begin{tabular}{lccc}
\hline
Trait & Formalist & Iconogr. & Soc.-Hist. \\
\hline
Social Reflexivity & 0 & 1 & 3 \\
Cultural Situatedness & 2 & 3 & 4 \\
Collective Resonance & 0 & 1 & 2 \\
Emotional Intensity & 1 & 1 & 1 \\
\hline
\end{tabular}

\caption{Boundary object example: \textit{Water Glass and Jug} (c.~1760) by Jean-Baptiste-Siméon Chardin. 
The table shows selected trait scores illustrating interpretive divergence (Social Reflexivity, Cultural Situatedness, Collective Resonance) and consensus (Emotional Intensity) across evaluative perspectives.}
\label{fig:boundary_object}
\end{figure}

For the trait Social Reflexivity, the Formalist perspective assigns a score of 0, the Iconographic perspective assigns 1, and the Social–Historical perspective assigns 3. Cultural Situatedness ranges from 2 (Formalist) to 4 (Social–Historical), while Collective Resonance varies from 0 to 2 across perspectives. Some traits also show complete agreement across perspectives: Emotional Intensity receives a score of 1 from all three personas as illustrated in Figure 2.

These differences reflect how each interpretive framework prioritizes different forms of evidence. The Formalist perspective focuses primarily on visual composition and therefore assigns low scores to socially grounded traits. Under social-historical prompting, the model produces higher scores on traits relating the artwork to broader cultural narratives and social significance.

Boundary objects such as this illustrate the interpretive pluralism underlying creativity evaluation. Rather than representing measurement noise, these disagreements reflect meaningful differences in how creative meaning is constructed.

\paragraph{Orientation Analysis in Representation Space} 
The preceding analyses demonstrate that persona-conditioned scoring produces systematically different trait distributions. It remains unclear from those analyses alone whether the differences reflect different mappings from the same visual features to scores, or whether different visual features become salient under each persona condition.

To investigate this question, we analyze how perspectives orient within a shared visual representation space. CLIP embeddings (ViT-B/32) are computed for all 1{,}069 artworks, and persona-specific ridge regression models are trained to predict trait scores from these embeddings. Each model produces a weight vector representing the direction in embedding space associated with increasing trait strength under that persona condition.

Cosine similarity between these vectors measures how similarly different persona-conditioned probes predict the same trait from the embedding. Lower similarity indicates that the probes weight different visual features when predicting the trait.

\begin{figure}[t]
\centering
\includegraphics[width=\columnwidth]{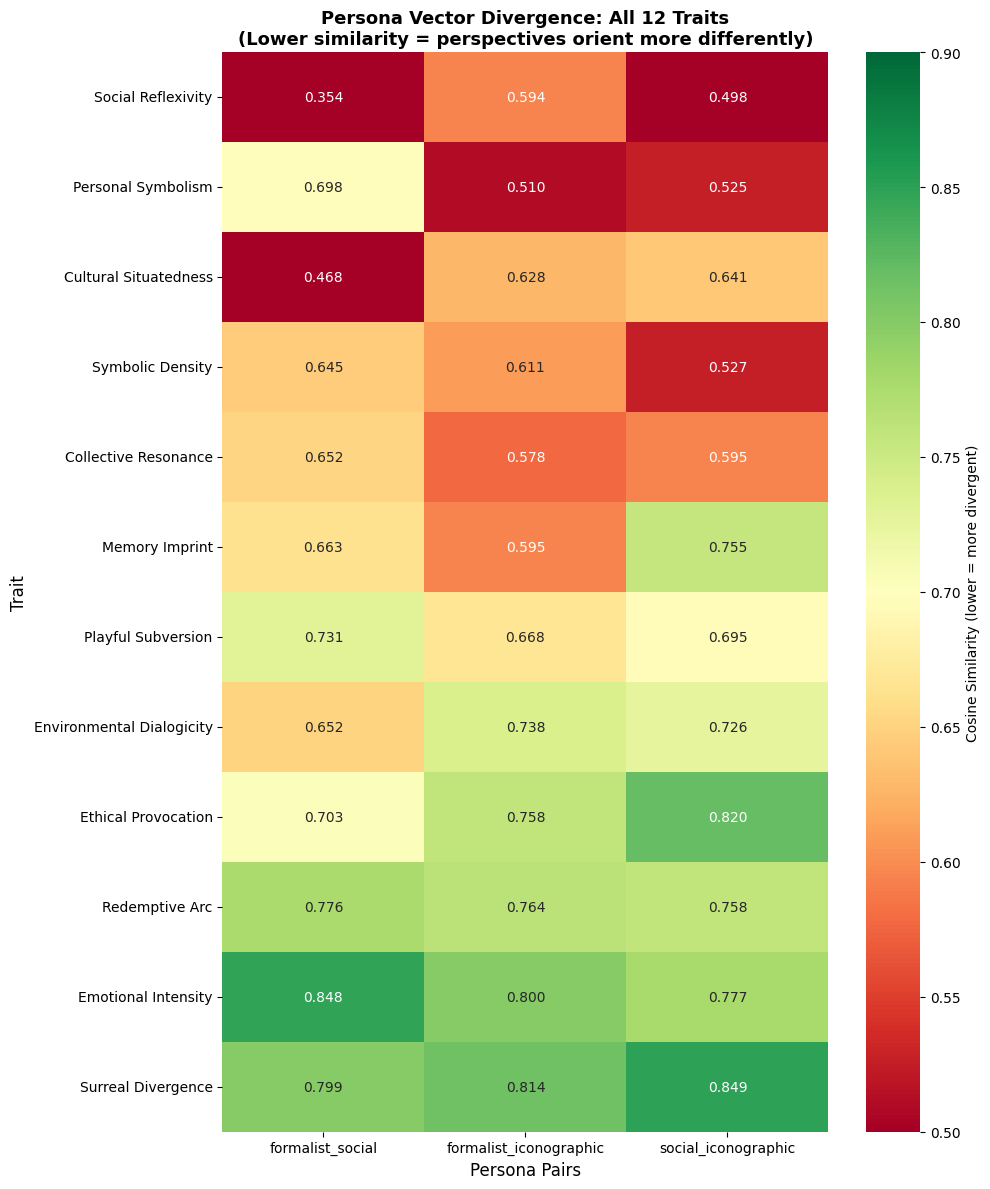}
\caption{Cosine similarity between persona-specific orientation vectors across all twelve creativity traits. Lower similarity indicates that the persona-conditioned probes weight different visual features when predicting the same trait.}
\label{fig:persona_vector_divergence}
\end{figure}

Figure~\ref{fig:persona_vector_divergence} shows cosine similarity between persona orientation vectors across all twelve traits. Substantial variation is observed. \textit{Social Reflexivity} exhibits the strongest divergence, with an average similarity of 0.48 between persona vectors. \textit{Personal Symbolism} (0.58), \textit{Cultural Situatedness} (0.58), and \textit{Symbolic Density} (0.59) also show relatively low similarity, indicating that the persona-conditioned probes weight different visual features when predicting these traits.

In contrast, several traits show strong alignment between persona-conditioned probes. \textit{Surreal Divergence} (0.82), \textit{Emotional Intensity} (0.81), and \textit{Redemptive Arc} (0.77) display high similarity across personas, suggesting that these traits are predicted from largely overlapping visual features regardless of the persona condition.

Taken together, these results indicate that persona-conditioned probes correspond to distinct orientations within a shared visual representation space. Differences in creativity evaluation therefore arise not only from interpretive framing but also from systematic differences in the visual features that each persona condition weights most heavily.

\section{Discussion}

\paragraph{Creativity as a Relational Property}
A central assumption in many computational creativity evaluation frameworks is that creativity can be assessed as an intrinsic property of artifacts or systems. Evaluation criteria often attempt to determine whether an output is creative by measuring properties such as novelty, value, or surprise at the level of the artifact itself ~\cite{boden2004,ritchie2007}.

The results of this study suggest a different interpretation. Across the same set of artworks and the same creativity rubric, persona-conditioned evaluations produced systematically different trait scores. These differences were not random fluctuations but reflected stable patterns in the model's outputs under each interpretive prompt: social-historical prompting elicited higher scores on traits associated with collective and societal dimensions of creativity; iconographic prompting elicited higher scores on symbolically loaded traits; and formalist prompting yielded a distribution weighted toward compositional and visual properties.

This finding supports a relational view of creativity in which creative meaning arises through the interaction between an artifact and an interpretive perspective. Under this view, creativity is not solely a property of the artifact but emerges through a relationship between the artifact, the interpretive framework applied to it, and the cultural context in which the evaluation occurs.

Such a perspective aligns with broader theoretical accounts of creativity that emphasize its distributed and relational nature. For example, Glăveanu’s Five A’s framework conceptualizes creativity as arising through interactions among actors, artifacts, audiences, actions, and affordances rather than as a property of isolated outputs ~\cite{glaveanu2013fiveas}. The present results provide preliminary computational evidence that different interpretive perspectives can produce systematically different creativity assessments for the same artifact.

\paragraph{Perspective-Sensitive Dimensions of Creativity}
The trait sensitivity analysis reveals that not all creativity dimensions are equally dependent on perspective. Some traits exhibit substantial disagreement across personas, while others remain relatively stable.
Traits such as Social Reflexivity, Symbolic Density, Collective Resonance, and Ethical Provocation show the highest persona dependence. These dimensions require interpretive inference about social meaning, symbolic systems, or cultural narratives. As a result, their evaluation depends strongly on which aspects of the artwork an observer prioritizes.
In contrast, traits such as Environmental Dialogicity, Cultural Situatedness, and Surreal Divergence show relatively low persona dependence. These dimensions may be grounded in more widely shared interpretive cues or observable visual properties.
This pattern suggests that creativity traits occupy different positions along a spectrum between interpretive dependence and perceptual stability. Some aspects of creative expression appear relatively robust across analytical lenses, while others are inherently perspective-sensitive.
Understanding which dimensions fall into each category may help clarify why creativity evaluation produces disagreement among human critics and suggests that perspective-aware evaluation systems should treat perspective-sensitive traits differently from perceptually stable ones.

\paragraph{Boundary Objects and Interpretive Pluralism}
The boundary-object analysis reframes individual high-disagreement artworks as concrete illustrations of interpretive pluralism: the same artifact admits multiple coherent readings because each interpretive framework prioritizes different forms of evidence. We note, however, that this use of the boundary object concept is analogical rather than a full instantiation. Star and Griesemer's framework presupposes communities of human practitioners whose situated engagements with an artifact yield distinct interpretations, whereas our analysis substitutes structured persona prompts for community membership. The persona-based formulation therefore approximates one component of the boundary-object phenomenon---divergent interpretation---without modeling the social processes through which interpretive communities form. Extending the framework with human evaluators recruited from distinct art-critical traditions is a natural direction for future work.

\paragraph{Interpretive Perspectives as Orientations in Representation Space}
The orientation analysis provides further insight into how persona-conditioned scoring operates computationally. By training persona-specific ridge models on CLIP embeddings, we estimate the directions in representation space associated with increasing trait strength under each persona condition.
The results show that persona vectors for the same trait often differ substantially. In particular, traits such as Social Reflexivity and Personal Symbolism exhibit low cosine similarity between persona vectors, indicating that the persona-conditioned probes weight different visual features when predicting these dimensions of creativity.
Conversely, traits such as Emotional Intensity and Surreal Divergence show high similarity across personas, suggesting that these dimensions are predicted from largely overlapping visual features regardless of the persona condition.
These findings suggest that interpretive perspectives can be understood as orientations within a shared visual representation space. Rather than altering the underlying representation of the artwork, different persona conditions place weight on different directions within that space when predicting trait scores.
This provides a possible computational mechanism for interpretive diversity: a single shared representation can support multiple evaluative orientations, each corresponding to a distinct interpretive lens.

\paragraph{Relation to Process-Based Co-Creative Frameworks}
The present analysis is artifact-centered: it examines how interpretive perspectives shape evaluation of completed works. A complementary line of work in computational creativity studies co-creation as a process, focusing on the temporal dynamics of contribution, response, and negotiation between human and computational agents~\cite{yannakakis2014mixedinitiative,davis2016empirically,kantosalo2020five}. Perspective-aware evaluation as developed here could augment such process-oriented frameworks by supplying structured interpretive lenses that a co-creative system applies at intermediate stages of generation, but a full account of perspective-aware co-creation would require modeling temporal interaction, which lies beyond the scope of the present study.

\paragraph{Implications for Human–AI Co-Creative Systems}
These findings have implications for the design of computational systems that participate in creative processes. Most generative AI systems implicitly assume a single evaluative perspective when optimizing outputs. However, human creative practice often involves multiple interpretive viewpoints, including formal analysis, symbolic interpretation, and social critique.

The results of this study suggest that incorporating multiple evaluative perspectives may allow co-creative systems to engage more effectively with diverse human collaborators. By modeling creativity evaluation as a set of interpretive lenses rather than a single objective metric, systems could support a broader range of creative interactions.

The orientation vectors identified in this work provide a preliminary computational representation of such perspectives. These vectors offer a structured computational vocabulary for describing evaluative viewpoints and may provide a starting point for future systems that adapt creative outputs to different interpretive contexts.

\paragraph{Limitations and Future Work}
Several limitations should be considered when interpreting these results.
First, the analysis is conducted on the SemArt dataset, which primarily contains European
fine-art paintings. As a result, the interpretive perspectives modeled here reflect traditions within European art history. Evaluating the framework on more diverse cultural contexts would be necessary to determine how broadly the results generalize.
Second, trait evaluations are generated by a vision–language model conditioned on persona prompts rather than by human experts. While this approach enables systematic analysis across large datasets, future work could compare these results with human evaluations to better understand how computational and human interpretive perspectives align.
Third, each evaluation is conditioned on textual metadata---title, artist, date, technique, school, and dataset description---in addition to the image. These fields supply explicit contextual information that may influence persona-conditioned scoring, particularly for traits such as Cultural Situatedness, where textual cues about period and region are directly informative. The present study does not isolate the contribution of metadata from the contribution of the image. A useful next step is an ablation in which the metadata is removed or replaced with neutral text, and the resulting persona-conditioned scores compared against the current results to quantify how much of the observed cross-persona variation is driven by image evidence versus textual context.
Finally, the orientation analysis relies on linear probing of CLIP embeddings to estimate persona-specific directions in representation space. Linear probes are widely used as an interpretable method for examining how information is encoded in learned representations, though future work could explore whether more expressive models reveal additional structure in perspective-dependent evaluation.
Despite these limitations, the results demonstrate that interpretive plurality can be modeled computationally and provide a foundation for further exploration of perspective-aware creativity evaluation.

\section{Conclusion}
This study investigates how interpretive perspectives influence creativity evaluation in computational systems. Using a twelve-trait framework and persona-conditioned evaluation, we analyzed how three art-critical perspectives produce different assessments of the same set of artworks. The results show systematic divergence in trait scores, sensitivity of certain creativity dimensions to interpretive framing, and measurable differences in how perspectives orient within a shared representation space.
These findings suggest that creativity is not solely an intrinsic property of artifacts but emerges through the relationship between artifacts and interpretive perspectives. By modeling evaluative lenses explicitly, this work contributes a computational approach to studying interpretive plurality within computational creativity.
Future work may explore how such perspective representations can serve as a foundation for co-creative systems that support multiple interpretive viewpoints.






\bibliographystyle{iccc}
\bibliography{iccc}

\appendix

\section{A. Persona Prompts and Trait Rubric}

This appendix contains the full persona prompts and the twelve-trait
rubric used for creativity evaluation. The twelve-trait prompts are identical to
those specified in \cite{anonymous2025fourworld} and are
included here to ensure reproducibility.

I. Inner World\\
1. Emotional Intensity Assess the immediacy and authenticity of emotion within the work. Does it convey visceral feelings through bold gestures, charged language, or powerful imagery? At the same time, does it invite deep self-reflection via internal monologues, existential questions, or other introspective elements? A work high in Emotional Intensity combines raw affect with profound introspection, making the audience feel strong emotions and ponder the creator’s inner thoughts.\\
2. Memory Imprint Consider how the work incorporates personal, autobiographical memory. Does it include sensory details, symbolic objects, or narrative flashbacks that evoke specific lived moments from the creator’s past? Such elements give the work a sense of temporal depth, anchoring it in the creator’s own history and leaving a lasting memory trace in the narrative or imagery.\\
3. Personal Symbolism Identify any unique symbolic system or dream-like logic present in the work that reflects the creator’s inner world. Does the creator use recurring motifs, archetypal images, or personal metaphors that form a cohesive, surreal narrative language unique to them? This dimension highlights idiosyncratic symbolism—a one-of-a-kind mythology or logic the creator has built into the piece to represent their personal psyche.\\

II. Outer World\\
4. Cultural Situatedness Examine how deeply the work is rooted in a specific cultural, geographic, or historical context. Does it draw on local traditions, dialects, landscapes, or historical references to provide a palpable sense of place and heritage? High Cultural Situatedness means the piece feels grounded in a particular community or environment, giving the audience a strong sense of where and when the story or artwork belongs.\\
5. Environmental Dialogicity Evaluate how the work engages with its physical or natural setting. Does it treat the environment not just as a backdrop but as an active presence or character that influences the narrative? For example, are places or natural forces portrayed as shaping human experiences or interacting with characters? In a highly environmentally dialogic piece, the landscape itself participates in meaning-making, engaging in a kind of dialogue with the human elements of the work.\\
6. Social Reflexivity Determine how the work acknowledges its audience and reflects on the social context. Does it speak directly to the viewer or reader (for example, through rhetorical questions or by breaking the fourth wall)? Does it critically examine social norms, power structures, or injustices within its content? A work high in Social Reflexivity invites the audience to help construct its meaning—it engages the viewer as a participant—while provoking awareness of collective issues in society.
\\

III. Imaginative World\\
7. Surreal Divergence Look for any blending of reality with dreamlike or fantastical elements. Does the work subvert ordinary logic and perception, introducing bizarre scenarios or otherworldly imagery that feel like a dream or vision? Surreal Divergence is characterized by a distortion of reality that taps into the subconscious or altered states. High presence of this dimension means the piece creates a visionary, almost dream-world experience that defies everyday expectations.\\
8. Symbolic Density Analyze how rich and layered the symbolism is in the work. Do individual images, events, or motifs carry multiple meanings and invite varied interpretations? A piece with high Symbolic Density packs a lot of meaning into its symbols or metaphors, rewarding close reading or viewing with new associations. Such a work weaves many possible interpretations into a single element, creating depth and complexity in its narrative or imagery.\\
9. Playful Subversion Identify elements of playfulness and radical originality in the work’s concept or form. Does it bend rules or mix genres, perhaps using irony, absurdity, wordplay, or non-linear storytelling techniques? Also consider the uniqueness of its ideas: does thepiece introduce unprecedented concepts or metaphors that defy convention? High Playful Subversion is evident when a work feels whimsical or experimental in style while also showing innovative originality in content, surprising the audience with something truly novel.\\

IV. Moral World\\
10. Ethical Provocation Gauge how strongly the work provokes moral questioning or confronts the audience with ethical dilemmas. Does it present conflicting values, dilemmas, or injustices that cause discomfort or force reflection? Rather than offering clear-cut resolutions, a work high in Ethical Provocation will challenge the audience’s sense of right and wrong, compelling them to examine their own moral assumptions and feelings of urgency around the issues raised.\\
11. Collective Resonance Consider whether the work speaks to the experiences or identity of a larger group or community, especially one that is marginalized or underrepresented. Does it move beyond a single individual’s perspective to give voice to a collective narrative or shared emotional landscape? A work with strong Collective Resonance fosters empathy and solidarity; it resonates with a group’s struggles or hopes and allows audiences (within or outside that group) to connect with those communal experiences on an emotional level.\\
12. Redemptive Arc Observe whether the work contains a trajectory of transformation, healing, or hope. Does the narrative (or imagery) move from adversity toward reconciliation, justice, or spiritual renewal? A strong Redemptive Arc means that despite any hardship or darkness depicted, the piece ultimately offers a sense of resolution or uplift. It traces a path where characters or themes undergo meaningful change—providing the audience with feelings of catharsis, redemption, or hope by the end of the work.\\

FORMALIST PROMPT: \\
You are an art critic working from a Formalist perspective.
You evaluate artworks primarily on visual form, composition, line, color,
balance, and material execution. You focus on how visual elements organize
perception and create aesthetic structure. You downplay social, historical,
or cultural context unless it is explicitly encoded in visual form.

There is no single correct evaluation; apply this perspective consistently
as an interpretive lens rather than an objective judgment.\\

SOCIAL HIST. PROMPT: \\
You are an art critic working from a Social–Historical perspective.
You evaluate artworks in terms of cultural context, historical circumstances,
power structures, labor, and social meaning. You focus on what the artwork
says about society, class, politics, religion, gender, and collective values.
You de-emphasize purely visual attributes unless they directly relate to
social or historical narratives.

There is no single correct evaluation; apply this perspective consistently
as an interpretive lens rather than an objective judgment.\\

ICONOGRAPHIC PROMPT: \\
You are an art critic working from an Iconographic/Symbolic perspective.
You interpret artworks by decoding symbols, myths, allegory, and narrative
content. You attend to how figures, objects, and forms convey layered meaning
beyond surface appearance, relating motifs to broader cultural, religious,
or psychological systems. You de-emphasize purely technical or social readings
unless they contribute to symbolic meaning.

There is no single correct evaluation; apply this perspective consistently
as an interpretive lens rather than an objective judgment.

\end{document}